\documentstyle[aps,preprint,eqsecnum]{revtex}
\begin{document}
\draft
\tighten
\title{The Goldberg--Kerr Approach to Lorentz Covariant Gravity}
\author{Brien C. Nolan\thanks{e-mail: nolanb@ccmail.dcu.ie}\\
School of Mathematical Sciences,\\
Dublin City University,\\ Glasnevin, Dublin 11}
\date{December 1995}
\maketitle
\begin{abstract}
In this paper, the approach to asymptotic electromagnetic fields introduced 
by Goldberg and Kerr ({\it J. Math. Phys.} {\bf 5} (1964)) is used to study
various aspects of Lorentz Covariant Gravity. Retarded multipole moments of
the source, the central objects of this study, are defined, and a sequence of conservation equations 
for these are derived
from the conservation of the energy-stress-momentum tensor of the source. 
The solution of the linearized Einstein field equation
is obtained in terms of the retarded moments for a general bound source field, 
correct to $O(r^{-4})$.
This is used to obtain the peeling--off of the linearized field, and
to study the geometric optics approximation for the field and for the 
energy-momentum of the field, given by the Landau--Lifschitz pseudotensor. 
It is shown that the energy-momentum 4-vector splits into the `total radiated
4-momentum' and the `bound 4-momentum of the source', similar to the case of
the electromagnetic field.
The role played by the conservation equations in studying the radiative
behaviour of the field is stressed throughout. 
In addition, in the case of a source which has 
only retarded pole, dipole and quadrupole moments, it is shown how to derive 
the arbitrary dependence of the field on a null coordinate. This allows 
comparison with the solutions for linearized 
gravity obtained by other authors.
\end{abstract}
\newcommand{\newc}{\newcommand}
\newc{\be}{\begin{eqnarray}}
\newc{\ee}{\end{eqnarray}}
\newc{\bgeq}{\be}
\newc{\eneq}{\ee}
\newc{\gams}{{\gamma^*}}
 
\def\sqr#1#2{{\vcenter{\hrule height.#2pt\hbox{\vrule width.#2pt                
height#1pt \kern#1pt \vrule width.#2pt}\hrule height.#2pt}}}                    
\def\square{\mathchoice\sqr75\sqr75\sqr{4.2}3\sqr{3.0}3}                        
                                                                                
\setbox1=\hbox{$\square$}
\def\dalemb{\raise 1pt\copy1}
\def\dalem{\mathop{\dalemb}\nolimits}

\newc{\bv}{b\cdot v}
\newc{\st}{{\cal{M}}}
\newc{\seq}{\stackrel{\ast}{{=}}}
\newc{\radt}{{}_{\rm rad}t}
\newc{\remt}{{}_{\rm rem}t}
\newc{\radp}{{}_{\rm rad}P}
\newc{\remp}{{}_{\rm rem}P}
\newc{\radq}{{}_{\rm rad}Q}
\newc{\remq}{{}_{\rm rem}Q}
\def\momp#1{^{#1}\!P}
\def\momq#1{^{#1}\!Q}

\newcommand{\zd}{\dot{z}}
\newcommand{\delt}{\nabla}
\newcommand{\kt}{\dot{k}}
\newcommand{\zdd}{\ddot{z}}
\newcommand{\ak}{a\!\cdot\! k}
\newcommand{\bk}{b\!\cdot\! k}
\newcommand{\ze}{\zeta}
\newcommand{\zeb}{\bar{\ze}}
\newcommand{\kz}[1]{ {\partial k^{#1}\over\partial\ze}}
\newcommand{\kzb}[1]{{\partial k^{#1}\over\partial\zeb}}
\newcommand{\dz}[1]{{\partial {#1}\over\partial\zeta}}
\newcommand{\dzb}[1]{{\partial {#1}\over\partial\zeb}}

\newc{\scri}{{\cal{J}}}
\newc{\btu}{\bigtriangleup}
\newc{\esm}{energy-stress-momentum }
\newcommand{\ldt}{\dot{l}}
\newcommand{\tp}{{\cal{T}}}
\newcommand{\mdt}{\dot{m}}
\newcommand{\rdt}{\dot{R}}
\newcommand{\tpdt}{{\dot{\tp}}}
\newc{\tpdtt}{\ddot{\tp}}
\newc{\tpdttt}{\stackrel{\dots}{\tp}}
\newcommand{\lam}{\lambda}
\newcommand{\fd}{\dot{F}}
\newc{\sbar}{\bar{s}}
\newc{\asb}{a\cdot\sbar}
\newc{\tdot}{\dot{T}}
\newc{\tdtt}{\ddot{T}}
\newc{\pdt}{\dot{p}}
\newc{\momterm}{\left[r^2\int K^{ijk}k_jv_k\,d\omega\right]}
\newc{\ck}{c\cdot k}
\newc{\go}{geometric optics }

\newcommand{\pdot}{\dot{p}}
\newcommand{\pdd}{\ddot{p}}
\newcommand{\ph}{\hat{P}}
\newcommand{\qh}{\hat{Q}}
\newcommand{\phd}{\dot{\ph}}
\newcommand{\phdd}{\ddot{\ph}}
\newcommand{\qht}{\dot{\qh}}
\newcommand{\qhtt}{\ddot{\qh}}
\newcommand{\qd}{\dot{Q}}
\newc{\nn}{\nonumber}
\section{introduction}
In previous articles \cite{HE,me}, 
the Goldberg--Kerr (GK) \cite{GK} approach to the
electromagnetic field was used to study various aspects of asymptotic
electromagnetic fields due to bounded sources. In particular, the geometric 
optics approximation and the arbitrary dependence of the field on a null
coordinate were established, and the relation of the multipole moments of the 
field to certain retarded moments of the source (integrals over null 
hyperplanes) was determined. One would like to be able to determine the same
relationship in the case of the gravitational field. However, the non-linear 
nature of Einstein's field equations prohibits this, and one must resort to 
approximation techniques to obtain results in this vein. Such techniques form
the keystone in the study of gravitational waves \cite{bladam,thorne}.

The simplest approximation is linearized General Relativity (GR), used in the
analysis of weak gravitional waves \cite{MTW,ll}. 
Here, the gravitational field is the 
Riemann tensor obtain from a first order perturbation to Minkowski 
space--time, quantified by writing the metric as 
\be g_{ij}=\eta_{ij}+\gamma_{ij}\,.\label{metric}\ee
Tensor indices are raised and lowered with the background Minkowski metric 
$\eta_{ij}$, and $\gamma_{ij}$ and its derivatives are taken to be small
of first order, so that any products among these terms which arise are 
neglected. From the metric (\ref{metric}) the linearized Riemann and Einstein 
tensors can be calculated, and the ensuing {\em linear} field equations
tackled, with or without a non-zero right hand side. Such discussions are 
usually augmented by a pseudotensor or effective stress tensor description
of the energy-momentum of the gravitational field.

The aim of this paper is to study aspects of linearized GR using the GK
approach to retarded fields due to bound sources. This involves obtaining the 
solution to the linearized field equations explicitly in terms of certian 
integrals over the source. In full GR, the linear approximation certainly 
does not hold all the way down to the source, but is usually applied in the 
distant wave zone. Hence, to distinguish these cases, the theory studied here
will be referred to as Lorentz Covariant Gravity (LCG). However, since most
of the results presented here are of an asymptotic nature, they will apply
to the distant wave zone of relativistic sources.

The fundamental object of this study is $\gamma_{ij}$, and the field equations 
that we consider may be written
\be \dalem\gams_{ij}&=&-16\pi T_{ij}\,,\label{feq}\\
{\gams^{ij}}_{,j}&=&0\,,\label{harm}\ee
where $\gams_{ij}$ is the trace reversed part of $\gamma_{ij}$, and the trace 
reversal of any symmetric second order tensor is
\be f^*_{ij}=f_{ij}-\frac{1}{2}\eta_{ij}f\,,\qquad f=\eta^{ij}f_{ij}\,.
\label{trrev}\ee
$\dalem$ is the d'Alembertian of flat space--time, and $T_{ij}$ is the 
energy-stress-momentum tensor of the source, which is conserved;
\be {T^{ij}}_{,j}=0\,.\label{esmcon}\ee
The gravitational field in this theory is given by
\be L_{ijkl}=\frac{1}{2}(\gamma_{il,jk}+\gamma_{jk,il}-\gamma_{ik,jl}
-\gamma_{jl,ik})\,,\label{lincur}\ee
where $\gamma_{ij}$ is obtained from the trace reversal of $\gams_{ij}$.

The field due to an isolated, extended, bound source will be considered. 
Thus the \esm tensor $T_{ij}$ of the source obeys the following
conditions \cite{GK}.

(1) There exists a time-like world line $C:x^i=z^i(\tau)$ with unit
tangent $v^i=dz^i/d\tau$, where $\tau$ is proper time along $C$.

(2) There exists a scalar function $h(\tau)$ such that $T^{ij}(x)=0$ for all
points $x$ such that
\be (x^i-z^i(\tau))(x_i-z_i(\tau))\geq{\rm max}
\left\{0,h^2(\tau)-[v_i(x^i-z^i(\tau))]^2\right\}\,.\label{bdsrc}\ee

Thus in an instantaneous rest frame (IRF) of $C$, $T^{ij}$ vanishes in the
infinite region extending to spatial infinity which is bounded by the future 
and past null cones with apex $z^i(\tau)\in C$ and the cylinder 
$|{\bf x}-{\bf z}(\tau)|=h(\tau)$. The conditions (1) and (2) above define the
world tube $W$ which is the support of $T^{ij}$, and in addition the 
weak assumption of `strong vanishing on the boundary of $W$' is made, 
that is it is 
assumed that $T^{ij}$ and a sufficient number of its derivatives vanish
on the boundary of $W$.

Using the world--line $C$, coordinates
and an associated null tetrad for Minkowski space-time ${\cal{M}}$ can be set 
up in the following way
(see \cite{teit,HE}). Let $x^a$ be any point of $\st$. 
The intersection of the past null cone $N$ with vertex $x^a$
and the world line $C$ is a unique point,
and defines a unique value $u$ of $\tau$, in terms of $x^a$,
which is referred to as retarded time. Tensor functions of $u$
defined on $C$ then become tensor fields
on $\st$ by parallel transport up the null cone $N$ to $x^a$;
e.g.~$v^a(x)\equiv v^a(\tau)|_{\tau=u}$. $a^a\,,b^a\,,c^a$ 
respectively are used to denote the second (acceleration), third and fourth
retarded time derivatives (denoted by a dot) of $z^a(u)$.
$k^a$ (null) is defined by
\bgeq x^a=z^a(u)+rk^a\,, \label{nuc} \eneq
where the normalization $k_av^a=-1$ is chosen.
Defining the stereographic coordinates 
(see appendix A of \cite{teit} for the
definition of the polar angles in this case)
\bgeq\zeta=e^{i\phi}\cot{\theta\over 2}\,,\label{stereo}\eneq
one can write
\bgeq k^a=p^{-1}(1+\ze\zeb,1-\ze\zeb,\ze+\zeb,-i(\ze-\zeb))\,,\label{kst}\eneq
where
\bgeq p=(1+\ze\zeb)v^0-(1-\ze\zeb)v^1-\ze(v^2-iv^3)-\zeb(v^2+iv^3)\,.
\label{stp}\eneq
One can also give
\bgeq {\partial\over\partial r}=k^a{\partial\over\partial x^a}\,,\qquad
{\partial\over\partial u}=v^a{\partial\over\partial x^a}+r\ak{\partial
\over\partial r}\,,\qquad {\partial\over\partial\zeta}=r{\partial k^a
\over\partial\zeta}{\partial\over\partial x^a}\,,\label{derivs}\eneq
\bgeq \eta^{ab}={1\over 2}p^2\left({\partial k^a\over\partial \zeta}
\kzb{b}+\kzb{a}\kz{a}\right)-k^av^b-v^ak^b+k^ak^b\,,\label{stmet}\eneq

\be{\partial\over\partial x^j}=-k_j{\partial\over\partial u}
-(v_j-(1+r\ak)k_j){\partial\over\partial r}+{p^2\over 2r}
\left(
{\partial k_j\over\partial\ze}{\partial\over\partial\zeb}
+{\partial k_j\over\partial\zeb}{\partial\over\partial\ze}
\right)\,,\label{ddx}\ee
and
\bgeq \dalem=\eta^{ab}{\partial^2\over\partial x^a\partial x^b}
={1\over r^2}\btu+(1+2r\ak)\left({\partial^2\over\partial r^2}
+{2\over r}{\partial\over\partial r}\right)-{2\over r}
{\partial\over\partial u}-2{\partial^2\over\partial u\partial r}\,,
\label{stdal}\eneq
where $\btu=p^2\partial^2/\partial\ze\partial\zeb$ is the Laplacian on the
unit sphere.

The retarded distance is
$r=-v_a(x^a-z^a(u))$, and the following
derivatives obtain:
\begin{mathletters}\label{dgeom}\begin{eqnarray}
u_{,a}&=&-k_a\,,\label{kdr}\\
v_{b,a}&=&-a_bk_a\,,\label{vd}\\
r_{,a}&=&-v_a+(1+r\ak)k_a\,,\label{rd}\\
k_{b,a}&=&{1\over r}
(\eta_{ab}+v_ak_b+k_av_b-(1+r\ak)k_ak_b)\,.
\label{kd}\end{eqnarray}\end{mathletters}
Here, $\ak=a_ak^a$.
From these, it is seen that $k^a$ is tangent to geodesics, with $r$
an affine parameter along its integral curves. One can use (\ref{kdr})
to differentiate tensor functions of $u$;
\bgeq {T^{b...}_{c...}}(u)_{,a}=-{\dot T}^{b...}_{c...}k_a\,.\eneq

The field equation (\ref{feq}) is solved by (see Synge \cite{sy}, p.407)
\be\gams_{ij}(x)=4\int_{N\cap W} T_{ij} d\sigma 
\seq 4\int\frac{T_{ij}(y^0,{\bf y})}{|{\bf x}-{\bf y}|} d_3y\,,
\label{retsol}\ee
where $d\sigma$ is the `absolute two-content of the three cell on the null
cone' (Synge, \cite{sy} p.429) which reduces to the familiar form in an
IRF of $C$. (Equations capped with an asterisk hold in the IRF, and boldface
letters represent spatial co-ordinates $x^\alpha,\,\alpha=1,2,3$.)
Events in $W$ are labelled $y^i=(y^0,{\bf y})$, and the integration 
is over the 
intersection of the past null cone $N$ with vertex $x^i$ with the world tube 
$W$.

In \S2, the basic variables of the study, the retarded multipole moments of 
the source are defined. A sequence of conservation equations for these 
moments are derived using (\ref{esmcon}), and their space-time derivatives
are calculated. The solution $\gams_{ij}$ is obtained to $O(r^{-4})$ for a
completely general source.

In \S3, the linearized curvature tensor is calculated. This allows
verification of Sachs' peeling--off theorem for linearized gravity
\cite{sachs}, but
for the more satisfactory case of a field due to an extended bounded source,
rather than one with its source confined to a time-like world line
\cite{dmc}. The role played by the conservation equations of \S2 in 
determining the Petrov type of the coefficients of the field is stressed,
particularly for the leading order (radiative) part of the field.
The results of this section are given in part in the Newman--Penrose
(NP) formalism.

The geometric optics approximation to different aspects of the field is
considered in \S4. First, it is demonstrated how the leading order term 
in the curvature tensor may be considered an approximate solution to the 
{\em vacuum} field equations which obeys the usual geometric conditions
of a radiation field (algebraic type N with shear-free twist-free geodesic 
rays). 
Second, the Landau--Lifschitz pseudotensor $t^{ij}$ is used to discuss the 
energy--momentum of the field. As in the case of the electromagnetic 
energy tensor \cite{HE,teit}, it is shown 
that the energy-momentum pseudotensor may be split into two parts, each
separately conserved, $_{\rm rad}t_{ij}$ and 
$_{\rm rem}t_{ij}=t_{ij}-{}_{\rm rad}t_{ij}$. The first term $\radt_{ij}$
has the form associated with a geometric optics field. From it, the 
`radiative 4-momentum' $\radp^i$ is constructed. This is associated with 
a `cloud of gravitons' which get radiated away  to ${\cal {J}}^+$, while
the corresponding $\remp^i$ is shown to determine a bound 4-momentum for the 
field \cite{teit}. 
In the course of this discussion the Sommerfeld outgoing radiation
conditions are discussed, and an interesting interplay between these and the
first of the series of conservation equations for the retarded moments of the  
source is pointed out.

As mentioned above,  it is possible to relate the multipole moments of the
field to the retarded moments of the source in the case of electromagnetism.
One expects the same to be true in the case of LCG, and in \S5 a decomposition
of the field, in the case of a pole-dipole-quadrupole source, is obtained.
This decomposition would allow one to obtain the stated relation, 
and also brings
out the expected arbitrary dependence of the field on a null coordinate
\cite{traut2}. The resulting alternative form for the solution $\gams_{ij}$
of (\ref{feq}) allows comparison with the solutions obtained by other authors
\cite{pir,sb}.
               
\section{Mathematical Preliminaries}                                         
                                                                                
In this section, the retarded multipole moments of the source         
$T_{ij}$ are defined, and the conservation equations which can be deduced from         
(\ref{esmcon}) are given. 
The space--time derivatives of the first           
few moments, which will be of use in the following sections, are derived.
                                                                                
The Goldberg--Kerr formalism is based on the intuitive notion that
as the field point moves infinitely far away from the source, the 
domain of integration in (\ref{retsol}) is asymptotically planar; the 
portion of the past null cone with apex $x^a$ which intersects $W$
flattens out.
Thus the central feature of the GK analysis is the use of retarded
multipole moments of the current 4--vector, defined by
\begin{mathletters}\label{mlp}
\bgeq \tp_{ij}=\int_P T_{ij}\, d\Sigma\,,\label{mlpo} \eneq
\bgeq \tp_{ij:k_1\cdots k_n}=\int_P T_{ij}\,\xi_{k_1}\cdots\xi_{k_n}
\,d\Sigma\,, \quad n=1,2,3,\dots\label{mlpn}\eneq                                    
\end{mathletters}
Eq.~(\ref{mlpo}) gives the retarded monopole moment, and (\ref{mlpn}) the           
retarded  $2^n$--pole moment.                                                   

Here, \bgeq\xi_i=y_i-z_i(u)\,,\eneq 
and on $P$,
\bgeq \xi^i=\left({
\bf \xi},{{\bf {k\cdot\xi}}\over k^0}\right)\seq
\left({\bf \xi},{\bf {k\cdot\xi}}\right)\,,\eneq
or equivalently,
\bgeq k_i\xi^i=0\,.\label{peq}\eneq
The domain of integration $P$ is
the null hyperplane which contains the null geodesic from the
field point $x$ to $C$, and $d\Sigma$ is the invariant volume element on
$P$, obeying
\bgeq d\Sigma\seq d_3y\,,\eneq
and it can be shown that
\bgeq d\Sigma_{,i}=\ak k_id\Sigma\,.\label{dsig}\eneq
Clearly, $\tp_{ij:k_1...k_n}$ is symmetric on all its indices after
the colon, and by its construction, obeys
\bgeq k^{l_1}\tp_{ij:l_1...l_n}=0\,.\label{kdmul}\eneq

Integrals of the form (\ref{mlpn}) arise naturally in this 
context in the following way. 
Points $y^i$ in the domain of integration of (\ref{retsol})
lie on the past null cone with vertex $x^i$, and so 
\begin{eqnarray} 
y^0&=&x^0-|{\bf {x-y}}|\nonumber\\
&=&z^0+rk^0-|{\bf {x-y}}|\nonumber\\
&\seq&z^0+{\bf {k\cdot\xi}}-\sigma\,,
\end{eqnarray}
where
\bgeq |{\bf {x-y}}|\seq r-{\bf {k\cdot\xi}}+\sigma\,,\eneq
and 
\bgeq \sigma = \sum_{n=1}^\infty b_n r^{-n}\,.\label{lsigdef}\eneq
The coefficients $b_n$ may be obtained from a power series expansion
of $ |{\bf {x-y}}|=|r{\bf k}-{\bf \xi}|$.
Then if $\sigma=0$, 
\begin{eqnarray}
y^0&\seq&z^0+{\bf {k\cdot\xi}}\nonumber\\
&=&z^0+{1\over k^0}{\bf {k\cdot \xi}}\label{yonp}\,,
\end{eqnarray}
so that according to (\ref{peq}), $y^i$ lies on $P$. 

Introducing the notation used in GK,  
\[ \left[F\right]=F(z^0+{1\over k^0}{\bf k\cdot\xi},{\bf y})\,,\]
it follows by applying the chain rule that
\bgeq \left[F\right]_{,a}=\left[F_{,a}\right]
+{k_a\over k^0}\left[F_{,0}\right]\,,\label{dfonp}\eneq
where $\left[F\right]_{,a}$ means `put
$y^0=z^0+{{\bf {k\cdot\xi}}\over k^0}$, and then differentiate
w.r.t. $y^a$' while $\left[F_{,a}\right]$ means `differentiate
w.r.t. $y^a$ and then put $y^0=z^0+{{\bf {k\cdot\xi}}\over k^0}$'.

The conservation equation for the source
(\ref{esmcon}) is used to obtain a sequence of conservation                      
equations for the retarded moments as follows. 
In (\ref{dfonp}), we take $F$ to be respectively          
$F=T^{ij}$ and $F=T^{ij}\xi^{k_1}\cdots\xi^{k_n}$, integrate over the
intersection of the null plane $P$ with the world tube $W$ and use              
(\ref{esmcon}) and the strong vanishing on the boundary to obtain               
\begin{mathletters}\label{econ}
\bgeq k^jD\tp_{ij}=0\,,\label{econo}\eneq
\bgeq k^jD\tp_{ij:k_1\dots k_n}+n\left\{\tp_{i(k_1:k_2\dots
k_n)}+k^j\tp_{ij:(k_2\dots k_n}v_{k_1)}\right\}=0\,.\label{econn}\eneq
\end{mathletters}                                                               

                                                                                
To obtain the derivatives of the retarded moments, the                
procedure given by Hogan and Ellis (HE) \cite{HE} is followed. 
The calculation for the monopole moment is given in full detail; 
derivatives of the higher moments          
are obtained in a similar way. 
From (\ref{mlpo}) and (\ref{dsig}), it follows that
\begin{eqnarray} \tp_{ij,k}&=&\int_P{\partial\over \partial
x^k}\left[T_{ij}\right]+\ak k_k\left[T_{ij}\right] d\Sigma\nonumber\\           
&=&\ak\tp_{ij}k_k+\int_P{\partial\over\partial x^k}\left[T_{ij}\right]
d\Sigma\,.\end{eqnarray}                                                        
The latter term is given by                                                     
\begin{eqnarray} \int_P {\partial\over\partial x^k}\left[T_{ij}\right]          
d\Sigma&=&\int_P\left[T_{ij,(1)}\right]{\partial y^0\over\partial x^k}          
d\Sigma \nonumber\\ 
&\seq&\int_P\left[T_{ij,(1)}\right]\left\{-k_k+{1\over
r}(\xi_k-({\bf {k\cdot\xi}})k_k)\right\} d_3y\,.\end{eqnarray}                   
To write this in terms of the retarded moments, note first that                
\[{\bf{k\cdot\xi}}\seq v^i\xi_i\,.\] 
In general,
\bgeq
\left[F_{,0}\right]=k^0\left\{{\partial\over\partial
u}\left[F\right]-\left[{\partial F\over\partial                                 
u}\right]\right\}\,,\label{duf}\eneq                                            
and using (\ref{dsig}), \bgeq \int_P{\partial\over\partial                      
u}\left[F\right] d\Sigma =D\int_P\left[F\right]                                 
d\Sigma\,.\label{duf1}\eneq 
Combining these results, one obtains
\bgeq\int_P{\partial\over\partial x^k}\left[T_{ij}\right]                       
=-k_kD\tp_{ij}+{1\over r}(D\tp_{ij:k}+v^lk_kD\tp_{ij:l})\,,\eneq 
and so
\begin{mathletters}\label{dts}
\bgeq\tp_{ij,k}=-\tpdt_{ij}k_k+{1\over
r}(D\tp_{ij:k}+v^lk_kD\tp_{ij:l})\,.\label{dto}\eneq
Similarly, it is found that                                                              
\bgeq \tp_{ij:k,l}=-\tpdt_{ij:k}k_l+{1\over
r}(v_k(\tp_{ij:l}+v^mk_l\tp_{ij:m})+
D\tp_{ij:kl}+v^mk_lD\tp_{ij:km})\,,\label{dt1}\eneq
and                                                                             
\begin{eqnarray} 
\tp_{ij:kl,m}&=&-\tpdt_{ij:kl}k_m+{1\over
r}\left(v_k(\tp_{ij:lm}+v^pk_m\tp_{ij:lp})+
v_l(\tp_{ij:km}+v^pk_m\tp_{ij:kp}) \right.\nonumber\\
&&\left.+D\tp_{ij:klm}+k_mv^pD\tp_{ij:klp}\right)\,.\label{dt2}
\end{eqnarray}
\end{mathletters}
                                                                                
                                                                                
Next, it is demonstrated how (\ref{retsol}) is evaluated. 
Again, the procedure of HE is followed.
A power series expansion at infinity in the variable $y^0$ yields
\[ T_{ij}(y^0,{\bf y})=\sum_{n=0}^\infty
{(-\sigma)^n\over n!}\left[ T_{ij,(n)}\right]\,,\]
where the subscript $(n)$ means differentiate $n$ times with respect to
$y^0$.
Writing
\bgeq H_n={(-\sigma)^n\over n!}{1\over |{\bf x}-{\bf y}|}\,,
\quad n=0,1,2,...,\label{hndef}\eneq
(\ref{retsol}) becomes (in the IRF)
\bgeq \gams_{ij}\seq
4\sum_{n=0}^\infty \int_P H_n\left[ T_{ij,(n)}\right] d_3y\,,\eneq
This can be written as a power          
series in $r^{-1}$, with the coefficients identified as IRF values of           
certain combinations of the retarded moments. For illustrative purposes,        
the calculation for the first two coefficients is given.
                                                                                
\newcommand{\thing}{(|{\bf \xi}|^2-({\bf {k\cdot\xi}})^2)}                      
With the $b_n$ defined by (\ref{lsigdef}), it is found that 
\bgeq b_1={1\over 2}         
(|{\bf \xi}|^2-({\bf {k\cdot\xi}})^2)                                           
\,,\eneq 
whence                                                                    
\begin{eqnarray} H_0&=&{1\over              
r}+{1\over r^2}{\bf{k\cdot\xi}}+O(r^{-3})\,,\\ H_1&=&-{1\over r^2}             
\thing+O(r^{-3})\,,\\ H_2&=&O(r^{-3})\,.\end{eqnarray}
\noindent Thus 
\begin{eqnarray} \gams_{ij}&=&{1\over                                        
r}\left\{4\int_P\left[T_{ij}\right] d_3y\right\} \nonumber\\ 
&&+{2\over r^2}\left\{ \int_P                                                              
2{\bf{k\cdot\xi}}\left[T_{ij}\right]-\thing\left[T_{ij,(1)}\right]             
d_3y\right\} +O(r^{-3})\,. \end{eqnarray}
The terms in the integrand here obey                                                                         
\begin{eqnarray} 
\int_P\left[T_{ij}\right]d_3y                                                
&\seq&\int_P\left[T_{ij}\right]d\Sigma =\tp_{ij}\,,\\
\int_P {\bf{k\cdot\xi}}\left[T_{ij}\right]d_3y                 
&\seq&-\int_P v^k\xi_k\left[T_{ij}\right]d\Sigma                               
=-\tp_{ij:k}v^k\,,\\
\thing&\seq&|{\bf{\xi}}|^2-(\xi^0)^2
=\xi_i\xi^i\,,\end{eqnarray} and so
\begin{eqnarray} \int_P \thing\left[T_{ij,(1)}\right]d_3y&\seq&                 
\int_P\left[T_{ij,(1)}\right] \xi^k\xi_k d_3y 
=D{\tp_{ij:k}}^k\,,\end{eqnarray}
where (\ref{duf}),                                                 
(\ref{duf1}) 
and  $\partial\left[ \xi^i\right]/\partial u=0$
have been used.
Then combining these results, one obtains
\bgeq \gams_{ij}=                                                               
{4\over r}\tp_{ij} -{2\over r^2}(2\tp_{ij:k}v^k+D{\tp_{ij:k}}^k)
+O(r^{-3})\,.\eneq                                                              

Using the same procedure, the next term in                                                   
the power series solution may be calculated to give                                               
\begin{eqnarray} \gams_{ij}&=&{4\over r}\tp_{ij}
-{2\over r^2}(2\tp_{ij:k}v^k+D{\tp_{ij:k}}^k)
+{2\over r^3}(2\tp_{ij:kl}v^kv^l-{\tp_{ij:k}}^k \nonumber\\
&&+2D{\tp_{ij:k}}^{kl}v_l 
+{\tp_{ij:k}}^{kl}Dv_l+{1\over 4}(D+\ak)D{\tp_{ij:kl}}^{kl}) +O(r^{-4})\,.
\label{soltor4} \end{eqnarray}                                                  
                                                                                
To end this section, it is pointed out that no assumptions, other than             
strong vanishing at the boundary, have been made about the source in            
order to obtain (\ref{soltor4}). In particular, the         
conservation equations (\ref{econ}) have not been imposed, 
nor has a restriction to a source         
possessing only a finite number of non--zero retarded multipole moments
been made.        
Finally, it should be pointed out that 
(\ref{esmcon}) is required to obtain the solution
(\ref{retsol}), so that the comments below regarding withholding the
conservation equations only strictly apply to quantities $\gams_{ij}$
formally defined by retarded integrals of the form (\ref{retsol}). This
will not affect the interpretation of the r{\^o}le played by the
conservation equation (\ref{econo}).

\section{Structure of the Curvature Tensor}
Having seen in the previous section how to evaluate the solution $\gams_{ij}$
of (\ref{feq}), the curvature tensor associated with this solution is now
evaluated and its structure examined. Sachs \cite{sachs} established the 
peeling--off theorem for linear fields with sources confined to a time-like 
world line; the same can now be done for the more realistic case of fields
due to extended bound sources (see \cite{dmc}). This is interesting in its own
right, but in addition will show up the important r{\^o}le played by the 
conservation equations in the determination of the peeling--off behaviour.

Thus the main object of this section is to calculate the linearized 
curvature tensor (\ref{lincur}) using the solution (\ref{soltor4}) 
correct to $O(r^{-4})$. Clearly, this would be an extremely lengthy
procedure, so in order to establish the main results with the minimum
fuss, the second derivatives of $\gams_{ij}$ are calculated and then
the NP Weyl tensor components are evaluated. This will give the following
limited peeling--off result;
\begin{mathletters}\label{peel}
\be 
 \Psi_4&=&{\Psi_4^{(0)}\over r}+O(r^{-2})\,,\\
 \Psi_3&=&{\Psi_3^{(0)}\over r^2}+O(r^{-3})\,,\\
 \Psi_2&=&{\Psi_2^{(0)}\over r^3}+O(r^{-4})\,,\\
 \Psi_1&=&O(r^{-4})\,,\\
 \Psi_0&=&O(r^{-4})\,,\ee
\end{mathletters}
where the $\Psi_A, A=0-4$ are calculated from the linearized 
curvature tensor (\ref{lincur}). See e.g. \cite{JN} for the definitions. 
These components will be evaluated on the null tetrad 
${\rm NT}=\left\{k^i, n^i, s^i, \sbar^i\right\}$
where
\be n_i=v_i-{1\over 2}k_i,\qquad 
s_i={1\over\sqrt{2}}p{\partial k_i\over \partial\ze},\qquad
\sbar_i={1\over\sqrt{2}}p{\partial k_i\over \partial\zeb}\,,
\ee
and so the only non-vanishing inner products are
\be k_in^i=-1,\qquad s_i\sbar^i=1\,.\ee

The formulae of \S 1 and the expressions above for the 
first derivates of the
retarded moments are used to obtain the following expressions.
(Note the use of (\ref{stmet}) in these calculations, and
$\pdt=-\ak p$, which follows immediately from (\ref{stp}).)
\be 
\tpdt_{ij,k}&=&-\tpdtt_{ij}k_k+{1\over r}(\delta^l_k+v^lk_k) 
D^2\tp_{ij:l}\,,\\
\tpdt_{ij:k,l}&=&-\tpdtt_{ij:k}k_l+{1\over r}(\delta^m_l+v^mk_l)
(v_kD\tp_{ij:m}+a_k\tp_{ij:m}+D^2\tp_{ij:km})\,,\\
\left.{\left.\tpdt_{ij:m}\right.}^m\right._{,l}&=&
-\left.\tpdtt_{ij:m}\right.^mk_l+{1\over r}(\delta^p_l+v^pk_l)
(2v^mD\tp_{ij:mp}+2a^m\tp_{ij:mp}+D^2{{\tp_{ij:m}}^m}_p)\,,\\
\left. \left.\tpdtt_{ij:m}\right. ^m\right._{,l}&=&
-{\tpdttt_{ij:m}}^mk_l+{1\over r}
(\delta^p_l+v^pk_l)
(2D^2(v^m\tp_{ij:mp})+D^3{{\tp_{ij:m}}^m}_p)\,.
\ee

Using these and the expressions for the first derivatives given in \S2,
the second derivative of the (trace reversed) metric perturbation can be
calculated. The resulting lengthy expressions are given in Appendix A.
Using these, the complete tensorial expression for $L_{ijkl}$ may be given,
correct to $O(r^{-4})$. However, the main results may be expressed more 
succinctly in the NP formalism.

Foremost of these results is the verification of (\ref{peel}), which 
is done using the expressions (\ref{first})-(\ref{last}) of Appendix A,
(\ref{lincur})
and the relation (\ref{trrev}) between $\gamma_{ij}$ and $\gams_{ij}$.
In the general case, it is found that
\be \Psi_4^{(0)}&=&-2(D+\ak)D\tp_{ij}s^i\sbar^j\,,\\
\Psi_3^{(0)}&=&
4\asb k^in^j\tp_{ij}+\asb \tp_i^i-2\sbar^i(a^j+\ak v^j)\tp_{ij}\nn\\
&&-2\sbar^i(D+\ak)D{\tp_{jm:}}^m+\sbar^j(D+\ak)D{\tp^i}_{i:j}\,.\ee
An examination of Appendix A will convince the reader that $\Psi_2^{(0)}$
is too long to merit inclusion. For simple illustrative purposes it may be
calculated for the case of a monopole particle, that is for a source for which
\[\tp_{ij}\neq 0,\qquad \tp_{ij:k_1...k_n}=0\,\quad n\geq1.\]
In this case, it is found that
\be \Psi_2 =-{m\over r^3}\,,\ee
where $m$ is a constant, which is the usual result.

It should be noted that in deriving eqs.(\ref{peel}) and the coefficients
above, the conservation equations (\ref{econ}) have been used repeatedly.
A more careful examination of the leading order term of the field
demonstrates what r{\^o}le these equations play. It is found that
\be L_{ijkl}= N_{ijkl}r^{-1}+O(r^{-2})\,,\label{ltor2}\ee
where 
\bgeq N_{ijkl}=N_{il}k_jk_k+N_{jk}k_ik_l -N_{ik}k_jk_l-N_{jl}k_ik_k\,,
\label{n1def}\eneq           
and 
\bgeq N_{ij}=(D+\ak)D\tp_{ij}-{1\over                                       
2}\eta_{ij}(D+\ak)D{\tp_m}^m\,.\label{n2def}\eneq                                            
Imposing the conservation equation (\ref{econo}), it is found that
\bgeq N_{ij}k^j=-{1\over 2}(D+\ak)D{\tp_m}^mk_i\,,\eneq 
and hence  
\bgeq N_{ijkl}k^l=0\,.\label{typen}\eneq                        
Thus, to leading order the            
field is Petrov type N, with degenerate principal null direction $k^i$.         
This vector has geodesic, twist--free, shear--free and expanding                
integral curves, the null rays of the radiation field. This will allow         
in \S4
a discussion of the geometric optics approximation to the field, by             
considering the field 
\bgeq {\cal{L}}_{ijkl}={N_{ijkl}\over r}\,,\eneq 
and         
working along the same lines as \S 4 of \cite{HE}.                              
                                                                                
It should be pointed out that                                                   
in general, i.e. without imposing the conservation equations                    
(\ref{econ}), the leading term in the field (\ref{ltor2}) is {\em not}          
type N. Indeed 
\[ N_{ijkl}k^l=D(k^lD\tp_{il})k_jk_k-D(k^lD\tp_{jl})k_ik_k\,,\] which will         
be non--zero in general. However, it is straightforward to show that
\bgeq N_{ijl[k}k_{m]}k^l=0\,,\label{type3}\eneq 
so that the leading order term is Petrov type         
III, with degenerate principal null direction $k^i$, which type of field        
is still characteristic of gravitational radiation. Thus it is seen how a 
fundamental property of the source, the monopole conservation equation, 
has consequences for the dynamics of the field. These consequences of
conservation will also manifest themselves when the flux of energy-momentum
of the field is considered.

\section{The Geometric Optics Approximation}
In this section, the geometric optics approximation to the Lorentz covariant
gravitational field is discussed. Two related aspects of this problem
are  considered. Firstly, it is shown that there exists a linearized curvature 
tensor which may be considered as an approximate solution to the field 
equations, and which has the geometric structure of a radiation field.
Secondly, the energy momentum of the full field is examined, using the
Landau--Lifschitz (LL) pseudotensor \cite{ll}. This is shown to split into
a geometric optis parts, which gets radiated away to $\scri^+$ along  null
geodesics, and a remainder term which in the way set out below, remains bound 
to the source.

The LL pseudotensor appropriate to linearized gravity is \cite{ll}
\be t^{ik}={1\over 16\pi}&&\left\{                                                  
2\Gamma^{nik}{\Gamma^p}_{np} -{\Gamma^{ni}}_p{\Gamma^{pk}}_n                    
-{\Gamma^{ni}}_n{\Gamma^{pk}}_p \right.\nonumber\\
&&-\left(2{\Gamma^{nm}}_m{\Gamma^p}_{np} -{\Gamma^{nm}}_p{\Gamma^p}_{mn}
-{\Gamma^{nm}}_n{\Gamma^p}_{mp}\right)\eta^{ik} \nonumber\\
&& +{\Gamma^{ki}}_p{\Gamma^{pn}}_n
+{\Gamma^{kn}}_n{\Gamma^{pi}}_p -{\Gamma^{km}}_p{\Gamma^{pi}}_m                 
-{\Gamma^{ki}}_m{\Gamma^{pm}}_p\nonumber\\
&& +{\Gamma^{ik}}_p{\Gamma^{pn}}_n
+{\Gamma^{in}}_n{\Gamma^{pk}}_p -{\Gamma^{im}}_p{\Gamma^{pk}}_m                 
-{\Gamma^{ik}}_m{\Gamma^{pm}}_p\nonumber\\
&& \left. +{\Gamma^{im}}_n{{\Gamma^k}_m}^n -{\Gamma^{im}}_m{\Gamma^{kn}}_n
\right\} \,,\label{pseudo}\ee
which is conserved,
\be {t^{ij}}_{,j}&=&0\,. \label{pstcon}\ee
Obviously, this is not constructed from the field $L_{ijkl}$, but from the
`field strengths', ${\Gamma^i}_{jk}$. So unlike the situation in 
electromagnetic theory, where the \go part of the energy--momentum tensor
of the field is the energy--momentum tensor of the \go part of the field,
the \go approximations must be defined separately. The relation between
the two will be clear. A different approach to the \go approxiation in 
linearized gravity involves a WKB analysis of the field equation (\ref{feq})
\cite{isaac}.

The first part of the discussion involves an analysis of the leading order 
term in the field,
\be {\cal{L}}_{ijkl}&=&{1\over r}N_{ijkl}\,,\label{lead}\ee
where $N_{ijkl}$ is defined in (\ref{n1def}) and (\ref{n2def}).

To show that (\ref{lead}) may be considered an approximate solution of
the linearized field equations, for large values of $r$, it must be 
shown to be (approximately) a linearized Riemann tensor with vanishing
Ricci tensor.
Away from the source, the linearized Ricci tensor constructed from (\ref{lead})
is
\be {\cal{R}}_{ij}&=&{{\cal{L}}_{ijk}}^j=0\,.\label{0ric}\ee
Now ${\cal{L}}_{ijkl}$ clearly has the symmetries of a Riemann tensor, so
it remains to show that it obeys (approximately) the 
linearized Bianchi identities,
\be L_{ij[kl,m]}&=&0\,. \label{lbid}\ee
Using some results given in \S3, one can calculate
\be N^*_{ij,k}&=&-2(D-\ak)(D+\ak)D\tp_{ij}k_k\nn\\
&&+{2\over r}\left\{(\delta^l_k+v^lk_k)(D+2\ak)(D+\ak)D\tp_{ij:l}
+3(a_k+\ak(v_k-k_k))\tpdt_{ij}\right.\nn\\
&&\left.+(b_k+\bv k_k+6\ak a_k+(\bk+6\ak^2)(v_k-k_k))\tp_{ij}\right\}\,.\ee
This is the central equation used to obtain 
\be {\cal{L}}_{ij[kl,m]}&=&{1\over r^2}C_{ij[kl}k_{m]}\,,\label{appbi}\ee
where
\be C_{ijkl}&=&8B_{[ij][kl]}\,,\\
 B_{ijkl}&=&\left\{(D+2\ak)(D+\ak)D\tp^*_{ik:l} 
+3(a_l+\ak v_l)\tpdt^*_{ik}\right.\nn\\
&&\left.+(b_l+6\ak a_l+(\bk+6\ak^2)v_l)\tp^*_{ik}\right\}k_j
+(D+\ak)D\tp^*_{ik}(\eta_{jl}+2k_jv_l)\,.\ee
A rest frame calculation along the lines of \S2 shows that
\be B_{ijkl}&\seq&k_j\int_P\left[ T^*_{ik,(3)}\zeta_l\right] d_3y
+(\eta_{jl}-2k_j\delta^4_l)\int_P\left[ T^*_{ik,(2)}\right] d_3y\,.
\label{binirf}\ee
The domain of integration here is the sphere
\[ |{\bf \ze}|\leq h(u)\,,\]
where $h$ is the `radius' of the source given in the introduction. Also,
since $T_{ij}(x)$ is, in each integral, compactly supported, there exist
scalars $M_n(u),\,\,n=0,1,2,\dots$ such that
\be |T^*_{ij,(n)}|\leq M_n(u)\,,\quad n=0,1,2,\dots\ee
Using this fact, the following bound can be obtained for (\ref{binirf});
\be |B_{ijkl}|\leq 4\pi h^3M_2+{4\over 3}\pi h^4 M_3\,,\label{bbound}\ee
which leads to a similar bound for $C_{ijkl}$.

Thus equations (\ref{0ric}), (\ref{appbi}) and (\ref{bbound}) show that 
${\cal{L}}_{ijkl}$ is an approximate solution of the vacuum linearized
Einstein equations. The accuracy of the approximation is determined by
the bounds $M_2$ and $M_3$.

It has already been shown that ${\cal{L}}_{ijkl}$ is Petrov type N, with
$k^i$ as repeated p.n.d. Therefore, the field picks out a unique null 
direction whose integral curves form a shear-free, twist-free and expanding
congruence of null geodesics. This ray geometry is characteristic of
a geometric optics field, and so ${\cal{L}}_{ijkl}$ is referred to as the
\go approximation to the linearized gravitational field. One usually thinks
of the field as `propagating in the direction $k^i$', and so it is natural
now to consider the question of energy-momentum transfer.

For this, the LL pseudotensor (\ref{pseudo}) is used. As pointed out above, 
this is not a functional of the field (curvature tensor), and so a 
seperate but related \go approximation is assumed. 
A detailed examination of the first two terms (coefficients of $r^{-2}$ and
$r^{-3}$) of $t^{ij}$ is required, to obtain which one can calculate
\be \gams_{ij,k}=-{4\over r}D\tp_{ij}k_k&+&{4\over r^2}
\left\{\tp_{ij}(v_k-k_k)+D\tp_{ij:k}\right.\nn\\
&&\left.+(v^l(2D+\ak)\tp_{ij:l}+a^l\tp_{ij:l}
+{1\over 2}(D+\ak)D{\tp_{ij:l}}^l)k_k\right\}+O(r^{-3})\,,\label{gamder}\ee
which leads to
\be \Gamma_{ljk}&=&-{2\over r}
(D\tp^*_{kl}k_j+D\tp^*_{jl}k_k-D\tp^*_{jk}k_l)\nn\\
&&+{1\over r^2}(D\tp^*_{kl:j}+D\tp^*_{jl:k}-D\tp^*_{jk:l}
+\tp^*_{kl}v_j+\tp^*_{jl}v_k-\tp^*_{jk}v_l
+a_{kl}k_j+a_{jl}k_k-a_{jk}k_l)\nn\\ 
&&+ O(r^{-3})\,,\label{gammasol}
\ee
where \nopagebreak[3]
\be a_{ij}&=&{1\over 2}(D+\ak)(D\left.{\tp^*_{ij:l}}\right.^l 
+v^l(2D+\ak)\tp^*_{ij:l}+a^l\tp^*_{ij:l}-\tp^*_{ij}\,.
\ee
Trace reversals are taken over indices before the 
colon. One then finds
\be a_{ij}k^j&=&-D{\tp_{il:}}^l-\tp_{il}v^l\nn\\
&&-
({1\over 4}(D+\ak)D{\tp^m_{m:l}}^l+v^l(D+{1\over 2}\ak)\tp^m_{m:l}
+{1\over 2}a^l\tp^m_{m:l}-{1\over 2}\tp^m_m)
k_i\,.\ee
Using this, it is found that
\be 
{\Gamma^{ki}}_i&=&{\gams^{ki}}_{,i}\equiv 0\,,\\
{\Gamma^k}_{ik}&=&{2\over r}D\tp^l_lk_j-{2\over r^2}
(D\tp^l_{l:j}+\tp^l_l v_j-a^l_lk_j)\,,\\
\Gamma_{ijk}k^k&=&{2\over r^2}(-\tp_{ji}+{1\over 2}\eta_{ji}+b_{[ji]})
+O(r^{-3})\,,\\
\Gamma_{ijk}k^i&=&{2\over r}D\tp^l_l k_jk_k - {2\over r^2}\left\{
\tp_{jk}+{1\over 2}\eta_{jk}\tp^l_l+{1\over 2}\tp^l_l(v_jk_k+v_kk_j)\right.
\nn\\
&&{1\over 2}D\tp^m_{m:j}k_k+{1\over 2}D\tp^m_{m:k}k_j
+(D{\tp_{kl:}}^l+\tp_{kl}v^l)k_j
+(D{\tp_{jl:}}^l+\tp_{jl}v^l)k_k\nn\\
&&\left.+({1\over 2}(D+\ak)D{\tp^m_{m:l}}^l+v^l(2D+\ak)\tp^m_{m:l}
+a^l\tp^m_{m:l}-\tp^m_m)k_jk_k\right\} +O(r^{-3})\,.
\ee
where
\begin{mathletters}\label{beqs}
\be b_{jl}&=&(\tp^m_m v_l+D\tp^m_{m:l}-2D{\tp_{lm:}}^m-2\tp_{lm}v^m)k_j\,,\ee
which gives
\be b_{jl}k^j&=&0\,\qquad b_{jl}k^l = \tp^m_m k_j\,.\ee
\end{mathletters}
It will be convenient to write
\be t^{ij}&=& 
\sum_{n=1}^\infty {_{(n)}t^{ij}\over r^{n+1}}\,,\ee
where
\be  {\partial _{(n)}t^{ij}\over \partial r}=0\,,\qquad n\geq 1\,,\ee
and to define
\be t^{ij}&=&\radt^{ij}+\remt^{ij}\,,\label{radrem}\ee
where
\be \radt_{ij}&=&{_{(1)}t_{ij}\over r^2}\,.\label{radt}\ee
The `rad' and `rem' here stand for `radiation' and `remainder' respectively.
Using (\ref{pseudo}) and (\ref{gammasol}) - (\ref{beqs}), it is found 
that
\be \radt^{ij}&=&{1\over 2\pi r^2}\left\{D\tp_{mn}D\tp^{mn}-{1\over 2}
(D\tp_m^m)^2\right\} k^ik^j\,,\label{rad}\ee
which is the usual form for the energy--momentum of a geometric optics or 
radiation field. It is straightforward to show that this obeys
\be {\radt^{ij}}_{,j}&=&0\,,\label{radcon}\ee
and since the total pseudotensor is conserved, the remainder is also 
conserved;
\be {\remt^{ij}}_{,j}&=&\left(
\sum_{n=2}^\infty {_{(n)}t^{ij}\over r^{n+1}}\right){}\!_{,j} =0\,.
\label{remcon}\ee
Now a long calculation using (\ref{pseudo}) and (\ref{gammasol}) - 
(\ref{beqs}) yields the following important result,
\be t^{ij}k_j&=&O(r^{-4})\,,\label{central}\ee
which from (\ref{rad}) is equivalent to
\be _{(2)}t^{ij}k_j&=&0\,.\label{notwo}\ee
Hogan and Ellis \label{HE} showed that the energy--momentum tensor of the
electromagnetic field due to a bound source obeys a set of equations 
similar to (\ref{rad}) - (\ref{notwo}) above. Thus a theorem 
of theirs can be applied to the linearized gravitational field:
\newtheorem{theorem}{Theorem}
\begin{theorem}
Let $K^{ijk}=U^{ik}k^j-U^{ij}k^k$, where
\[ U^{ik}={1\over r^2} _{(2)}t^{ik}+\sum_{n=3}^\infty{1\over r^n}\left\{
{ _{(n)}t^{ik}\over n-1}-{ _{(n)}t^{il}k_lv^k\over (n-1)(n-2)}\right\}\,.\]
Then one can write
\be {K^{ijk}}_{,k}& =& \sum_{n=2}^\infty {_{(n)}t^{ij}\over r^{n+1}} 
=\remt^{ij}\,.\label{theoremres}\ee
\end{theorem}
{\em Proof}: See \cite{HE}, p.202.

To examine the transfer of energy--momentum by the field, one considers the
flux of 4--momentum across the following fundamental surfaces. This follows
exactly the procedure of HE for the electromagnetic case.

Consider the 4-volume $V$ of Minkowskian space time bounded by two time--like 
3-surfaces $r=r_1$ and $r=r_2>r_1$ and by the null cones $u=u_1$ and 
$u=u_2>u_1$. $r_1$ is taken to be large enough so that we are outside the
source, and so away from the world-line $C$ given by $r=0$. Thus $t^{ij}$
is non-singular in $V$. The fluxes of 4-momentum across the boundaries of $V$
(in the directions of increasing $r,u$ on the appropriate surfaces) are given
by
\be \momp{(A)}^i&=&r_A^2\int_{u_1}^{u_2}\, du \int t^{ij} r_{,j}\, d\omega\,,
\label{pflux}\ee
across $r=r_A, (A=1,2)$ and 
\be \momq{(A)}^i &=&-\int_{r_1}^{r_2}r^2\,dr\int t^{ij}k_j\,
d\omega\,,\label{qflux}
\ee
across $u=u_A$. In these integrals, $d\omega$ is the area element on the unit
2-sphere. The conservation equation ${t^{ij}}_{,j}=0$ gives the conservation 
law
\be \momp{(1)}^i+ \momq{(1)}^i&=&\momp{(2)}^i+ 
\momq{(2)}^i\,,\label{momcon}\ee
which says that the total 4-momentum entering $V$ is equal to the total
4-momentum leaving $V$.

Using the decomposition of the energy-momentum pseudotensor (\ref{radrem}),
one can write
\be P^i&=&\radp^i+\remp^i\,,\label{pradrem}\ee
on $r=$constant, and
\be Q^i&=&\radq^i+\remq^i\,,\label{qradrem}\ee
on $u=$constant, where the first terms on the right hand sides are obtained 
by using $\radt^{ij}$ in the definitions
(\ref{pflux}) and (\ref{qflux}), and the second terms on the right hand
side by using $\remt^{ij}$.

From (\ref{rad}), one finds that
\be \radp^i&=&{1\over 2\pi}\int_{u_1}^{u_2}\int 
(D\tp_{mn}D\tp^{mn}-{1\over 2}(D\tp_m^m)^2)k^i\,d\omega\,,\label{radflux}\ee
and
\be \radq^i&=&0\,.\ee
Noticing that $\radp^i$ is independent of $r$, 
these results give trivially the conservation law
\be _{\rm rad}\momp{(1)}^i+ _{\rm rad}\momq{(1)}^i&=&
_{\rm rad}\momp{(2)}^i+ _{\rm rad}\momq{(2)}^i\,,\ee
which also follows from (\ref{radcon}).
This also means that $\radp^i$  may be used to calculate the flux of
4-momentum at $\scri^+$ ($r\rightarrow \infty,\,u$ finite). Indeed, since 
$k^i$ is future pointing and the coefficient
\[ D\tp_{mn}D\tp^{mn}-{1\over 2}(D\tp_m^m)^2 \geq 0\]
(see below), one sees that $\radp^i$ is future pointing, and so this vector
is referred to as the {\em total radiated 4-momentum of the field} across
$r=$ constant in the retarded time $u_2-u_1$.

At this stage, the question of outgoing radiation conditions may be addressed. 
Trautman has expressed Sommerfeld's outgoing radiation conditions               
as follows \cite{traut1,pete74}: there exist coordinate systems and a           
tensor $f_{ij}=O(r^{-1})$ such that                                             
\begin{mathletters}\label{out}
\begin{eqnarray}
g_{ij}&=&\eta_{ij}+O(r^{-1})\,,\label{out1}\\
g_{ij,k}&=&f_{ij}k_k+O(r^{-2})\,,\label{out2}\\
f^*_{ij}k^j&=&O(r^{-2})\,,\label{out3}
\end{eqnarray}
\end{mathletters}
with $f^*_{ij}$ obtained from (\ref{trrev}).                                    
                                                                                
The energy--momentum pseudotensor calculated from                               
(\ref{out}) turns out to obey \bgeq t^{ij}={1\over 32\pi}\left\{                
f^*_{mn}{f^*}^{mn}-{1\over 2}({f^*}_m^m)^2 \right\} k^ik^j                    
+O(r^{-3})\,,\label{llout}\eneq                                 
and the         
coefficient of $k^ik^j$ here is positive.        
Indeed, given any symmetric second order tensor $A_{ij}$ on $\st$ obeying 
$A_{ij}k^j=0$, one can write 
%
%
%
\begin{eqnarray}                                                                
A_{ij}&=& \alpha k_ik_j+\beta (k_is_j+k_js_i)+                                  
{\bar\beta}(k_i\sbar_j+k_j\sbar_i)\nonumber\\ && +\gamma                            
s_is_j+{\bar\gamma}\sbar_i\sbar_j +\delta(s_i\sbar_j+s_j\sbar_i)\,,                     
\nonumber\end{eqnarray}                                                         
where $\alpha$ and $\delta$ are real valued and the null tetrad of 
\S 3 has been used. Then                               
$ A_i^i=2\delta$ and $ A_{ij}A^{ij}=2\gamma{\bar\gamma}$                 
so that                                                                         
\[ A_{ij}A^{ij}-{1\over 2}(A_i^i)^2=                                               
2\gamma{\bar\gamma}\,, \] which is positive.        
This is                                                                         
applied to $f^*_{ij}$, using (\ref{out3}), to show that coefficient
of $k^i$ in the integrand in (\ref{radflux}) is positive, 
so that the integral leads to a {\em positive} 
outward flux of 4--momentum of the field.                                                                      

Now from (\ref{gamder}), it is seen that                                              
(\ref{out1}) and (\ref{out2}) will be satisfied for the field by taking         
\be                                                                              
f_{ij}=-(4D\tp_{ij}-2\eta_{ij}D{\tp_k}^k)r^{-1}\,.\ee
Then the key equation (\ref{out3}) which is needed for
the procedure above is                                                              
\be k^jD\tp_{ij}=0\,,\ee which is exactly (\ref{econo}), the                              
monopole conservation equation of \S 2.                                                 

This may be compared with the result obtained by Hogan for a moving monopole
particle \cite{pete74}. Consider then a source for which 
\be \tp_{ij}\neq 0\,,\qquad \tp_{ij:k_1\dots k_n}=0\,, n\geq1\,.\ee
From (\ref{econn}) with $n=1$, one obtains 
\be \tp_{ij}=mv_iv_j\,,\ee for some scalar $m$. Then the outgoing radiation
conditions, i.e. (\ref{usecon1}) give
\be \left(\tp_{ij}k^j\right)^\cdot = -\mdt v_i-ma_i=0\,,\ee
from which (contract with $v^i$, then $a^i$, using $v_ia^i=0$) it is found
$\mdt=0$ and $a^i=0$. This is exactly the result obtained by Hogan; the 
outgoing radiation conditions force the acceleration of the particle to be 
zero. What is seen here is how this relates to the conservation of energy
momentum of the source, i.e. of the particle.
In general, this shows how a fundamental property of the source, its 
conservation, has important consequences for the dynamics of the field. 

Next consider the remainder terms in (\ref{pradrem}) and (\ref{qradrem}).
Using (\ref{theoremres}) and Stokes' theorem, one can write
\be \remp^i&=&\momterm_{u=u_2}-\momterm_{u=u_1}\,,\label{prem}\ee 
and
\be \remq^i&=&-\momterm_{r=r_2}+\momterm_{r=r_1}\,.\label{qrem}\ee
Using the definitions of the theorem and (\ref{central}), one can write
\be r^2\int K^{ijk}k_jv_k\,d\omega&=&\sum_{n=3}^\infty 
{f^i_n(u)\over r^{n-2}(n-2)}\,,\ee
where
\be f^i_n(u)&=&\int{} _{(n)}\!t^{ij}k_j\,d\omega\,,\quad n=3,4,5,\dots\ee
This allows one to write
\begin{mathletters}\label{rempflux}
\be 
_{{\rm {rem}}}\momp{(1)}^i&=& 
\sum_{n=3}^\infty 
{ f^i_n(u_2)-f^i_n(u_1) \over r_1^{n-2}(n-2) }\,,\\
_{{\rm {rem}}}\momp{(2)}^i&=& \sum_{n=3}^\infty 
{ f^i_n(u_2)-f^i_n(u_1) \over r_2^{n-2}(n-2) }\,,
\ee
\end{mathletters}
and
\begin{mathletters}\label{remqflux}
\be 
_{{\rm {rem}}}\momq{(1)}^i&=& \sum_{n=3}^\infty
{f^i_n(u_1)\over (n-2)} \left( {1\over r_1^{n-2}} - {1\over r_2^{n-2}}
\right)\,,\\
_{{\rm {rem}}}\momq{(2)}^i&=& \sum_{n=3}^\infty
{f^i_n(u_2)\over (n-2)} \left( {1\over r_1^{n-2}} - {1\over r_2^{n-2}}
\right)\,,\ee
\end{mathletters}
which lead to the expected conservation law,
\be _{{\rm {rem}}}\momp{(1)}^i+_{{\rm {rem}}}\momq{(1)}^i&=&
_{{\rm {rem}}}\momp{(2)}^i+_{{\rm {rem}}}\momq{(2)}^i\,.\ee
Unlike the case for $\radp^i$, the fluxes in (\ref{rempflux}) depend on $r$.
Eq. (\ref{remqflux}) indicates a leakage of 4-momentum across null cones, 
so that the remainder 4-momentum is not purely radiative.

One can make the same observations which Hogan and Ellis made for the 
electromagnetic field. The total flux of 4-momentum across $r=$ constant
in a proper time $u_2-u_1$ is
\be \remp^i&=& p^i(u_2)-p^i(u_1)\,,\label{particle}\ee
with
\be p^i(u)&=&
\sum_{n=3}^\infty 
{ f^i_n(u) \over r_1^{n-2}(n-2) }\,.\ee
This vanishes in the limit $r\rightarrow\infty$ and so is non-radiative in 
character (no flux at $\scri^+$). This qualifies the characterization of 
$\radp^i$ as being the {\em total} radiated 4-momentum. 
Eq.(\ref{particle}) may be said to describe `particle-like behaviour',
in that the flux of 4-momentum between $u=u_1$ and $u=u_2$ is simply the
difference in the values of the `4-momentum ' $p^i(u)$ at the two times.
For this reason we refer to $p^i(u)$ as the bound 4-momentum of the source
\cite{HE,teit}. Thus Teitelboim's idea of the bound 4-momentum of the 
electromagnetic field \cite{teit1} may be generalised to 
(at least) the linearized gravitational field.

The splitting of the flux of 4-momentum across $r=$ constant surfaces defines 
the geometric optics approximation for the energy--momentum of the field.
The relation to the geometric optics part of the field (curvature tensor) 
itself is clear; only terms which are involved in the construction of 
${\cal{L}}_{ijkl}$ are involved in the construction of $\radt^{ij}$.
Thus associated with ${\cal{L}}_{ijkl}$ is the total radiated 4-momentum,
which detaches itself from the source and radiates away to $\scri^+$ at the
speed of light. The energy-momentum which is left behind remains bound to the
source.
\section {Arbitrary Dependence on a Null Coordinate}

In this section, results are obtained which can be used to determine the 
relation between the retarded multipole moments of the source and the 
multipole moments of the field. 
This will involve determining a decomposition of the field into 
arbitrary functions of $u$ (which is a null coordinate), and known functions
of angle. As in the electromagnetic case, radiative linearized gravitational
fields must contain arbitrary functions of a null coordinate in order to
convey information \cite{traut2}. The decomposition obtained will allow
a rewriting of the solution $\gamma_{ij}$ in a form which allows direct
comparison with  other approaches to linearized gravity \cite{sb,pir}.

In order to keep the calculations to a reasonable length, 
attention is restricted 
to a source for which only the retarded pole, dipole and 
quadrupole moments are non--zero. Such a source will suffice
to display generic behaviour. Thus 
\bgeq \tp_{ij:k_1\dots k_n}=0\,,\quad n\geq3\,.\label{gpdq}\eneq

The decomposition of the retarded  pole, dipole and quadrupole
moments is obtained by analysing the conservation equations (\ref{econ}).

Defining the tensors
\bgeq m_i=\tp_{ij}k^j\,,\qquad m_{i:k_1\dots
k_n}=k^j\tp_{ij:k_1\dots k_n}\,,\label{mmu}\eneq 
the first few conservation laws may be rewritten as                                                                              
\bgeq \mdt_i=0\,,\label{usecon1}\eneq                                           
\bgeq \mdt_{i:j}+\tp_{ij}+m_iv_j=0\,,\label{usecon2}\eneq
\bgeq \mdt_{i:jk}+\tp_{ij:k}+\tp_{ik:j}+m_{i:j}v_k+m_{i:k}v_j=0\,,
\label{usecon3} \eneq                                                           
\bgeq \mdt_{i:jkl}+\tp_{ij:kl}+\tp_{ik:jl}+\tp_{il:jk}
+m_{i:jk}v_l+m_{i:kl}v_j+m_{i:lj}v_k=0\,.\label{usecon4}\eneq
                                                   
The first of these may be rewritten as
\bgeq \tp_{ij}=-{1\over 2}(\mdt_{i:j}+\mdt_{j:i}) -{1\over 2}(m_iv_j+m_jv_i)
\,.\label{pdcm}\eneq
From (\ref{usecon2}), one can deduce that there exists a tensor $Q_{ijk}$
such that
\bgeq \tp_{ij:k}=-m_{i:j}v_k-{1\over 2}\mdt_{i:jk}-Q_{ijk}\,, \label{ddcm}
\eneq
where $Q_{ijk}$ obeys
\bgeq Q_{ijk}=-Q_{ikj}\,,\label{q3eq1}\eneq
\bgeq Q_{ijk}k^k=-Q_{ikj}k^k=m_{i:j}\,.\label{q3eq2}\eneq
Similarly, using (\ref{usecon3}) along with (\ref{gpdq}), one finds that there
exists a tensor $Q_{ijkl}$ such that
\bgeq \tp_{ij:kl}=-{1\over 2}(v_km_{j:il}+v_lm_{j:ik})+Q_{jikl}\,,
\label{qdcm}\eneq
and $Q_{ijkl}$ obeys
\bgeq Q_{ijkl}=Q_{ijlk}\,,\qquad Q_{ijkl}+Q_{iklj}+Q_{iljk}=0\,,
\label{q4eq1}\eneq
\bgeq Q_{ijkl}k^l=-{1\over 2}m_{i:jk}\,,\qquad Q_{ijkl}k^j=m_{i:kl}\,,
\label{q4eq2} \eneq
\bgeq Q_{ijkl}k^jk^k=Q_{ijkl}k^jk^l=Q_{ijkl}k^kk^l=0\,.\label{q4eq3}
\eneq 

The procedure runs as follows: using the formulas for the derivatives of the 
retarded moments, 
$\zeta$--derivatives of $Q_{ijk}$ and $Q_{ijkl}$ are obtained. 
These give a pair
of first order simultaneous differential equations, which are in an
integrable form. The `constants of integration' yield the required
arbitrary functions of $u$.

To begin, recall that
\[{\partial\over\partial\zeta} =r\dz{k^i}{\partial\over\partial x^i}\,.\]
Then using the derivatives (\ref{dts}) one finds
\bgeq \dz{m_{i:jk}}=-(\tp_{ij:kl}+\tp_{ik:jl})\dz{k^l}\,,\eneq
and so
\bgeq \dz{\mdt_{i:jk}}=-(D\tp_{ij:kl}+D\tp_{ik:jl})\dz{k^l}\,.\eneq
Using (\ref{dt1}) and (\ref{dts}) one obtains
respectively
\bgeq \dz{\tp_{ij:k}}=(v_k\tp_{ij:l}+D\tp_{ij:kl})\dz{k^l}\,,\eneq
and
\bgeq \dz{m_{i:j}}=-\tp_{ij:l}\dz{k^l}\,.\eneq
Using these derivatives and the definition (\ref{ddcm}) of $Q_{ijk}$
gives 
\bgeq \dz{Q_{ijk}}=-{1\over 2}{\partial\over\partial u}
\left\{(\tp_{ij:kl}-\tp_{ik:jl})\dz{k^l}\right\} \,.\label{q3zeta}\eneq
Similarly, one can use (\ref{dt2}) and (\ref{dts}) in the definition 
(\ref{qdcm}) (remember $v^i=v^i(u)$) to obtain
\bgeq \dz{Q_{jikl}}={1\over 2}\left\{ (\tp_{ji:lm}-\tp_{jl:im})v_k 
+(\tp_{ji:km}-\tp_{jk:im})v_l\right\} \dz{k^m}\,.\label{q4zeta}\eneq
Now from (\ref{q3zeta}) one can deduce the existence of tensors
$R_{ijk}$ and $\qh_{ijk}(u)$, both having the same symmetries as $Q_{ijk}$,
and $\qh_{ijk}$ depending only on $u$, such that
\bgeq Q_{ijk}=\rdt_{ijk}+\qh_{ijk}\,, \label{q3sol}\eneq
and
\bgeq \dz{R_{ijk}}=-{1\over 2}(\tp_{ij:km}-\tp_{ik:jm}) \dz{k^m}\,.
\label{rzeta}\eneq
Comparing this with (\ref{q4zeta}), one can write
\bgeq Q_{jikl}=-R_{jil}v_k-R_{jik}v_l+\qh_{jikl}(u)\,, \label{q4sol}\eneq
where $\qh_{jikl}$ has the same symmetries as $Q_{jikl}$ and depends only
on $u$.

Eq.(\ref{rzeta}) is solved as follows. Using (\ref{q4sol}) and the first of 
(\ref{q4eq2}), 
\bgeq m_{j:ik}=-2R_{jik}+2v_kR_{jil}k^l-2\qh_{jikl}k^l\,,\eneq
which on using (\ref{q4sol}) and (\ref{qdcm}) gives
\bgeq \tp_{ij:kl}= -2R_{ijm}k^mv_kv_l+\qh_{ijkm}k^mv_l+\qh_{ijlm}k^mv_k 
+\qh_{ijkl}\,. \eneq
Then
\begin{eqnarray}(\tp_{ij:kl}-\tp_{ik:jl})\dz{k^l}&=&
(\qh_{ijkl}+\qh_{ijlm}k^mv_k-\qh_{ikjl}-\qh_{iklm}k^mv_j) \dz{k^l}
\nonumber\\
&=&{\partial\over\partial\ze}\left(
\qh_{ijlm}(\delta^m_k+{1\over 2}k^mv_k)k^l
-\qh_{iklm}(\delta^m_j+{1\over 2}k^mv_j)k^l\right)\,.
\end{eqnarray}
Comparing this with (\ref{rzeta}), one can integrate and give
\bgeq R_{ijk}=-{1\over 2}\left\{
\qh_{ijlm}(\delta^m_k+{1\over 2}k^mv_k)k^l
-\qh_{iklm}(\delta^m_j+{1\over 2}k^mv_j)k^l\right\}\,. \label{rsol}\eneq
The constant of integration which arises here can be absorbed into 
$\qh_{ijk}$ and $\qh_{ijkl}$, and can thus be set equal to zero without
changing $Q_{ijk}$ or $Q_{ijkl}$. From (\ref{q3sol}) and (\ref{q4sol}),
\bgeq Q_{ijk}=-{1\over 2}{\partial\over\partial u} \left\{
\qh_{ijlm}(\delta^m_k+{1\over 2}k^mv_k)k^l
-\qh_{iklm}(\delta^m_j+{1\over 2}k^mv_j)k^l\right\}+\qh_{ijk}\,,
\label{q3}\eneq
and
\begin{eqnarray}
Q_{jikl}&=&
{1\over 2}v_l
\left\{ \qh_{jipm}(\delta^m_k+{1\over 2}k^mv_k)k^p
-\qh_{jkpm}(\delta^m_l+{1\over 2}k^mv_l)k^p \right\}\nonumber\\
&& +{1\over 2}v_k
\left\{ \qh_{jipm}(\delta^m_l+{1\over 2}k^mv_l)k^p
-\qh_{jlpm}(\delta^m_i+{1\over 2}k^mv_i)k^p \right\}\nonumber\\
&&+\qh_{jikl}(u)\,.\label{q4}\end{eqnarray}

The tensors $\qh_{ijk}$ and $\qh_{ijkl}$ will be specified by the
source via these last two equations and (\ref{ddcm}) and (\ref{qdcm}). 
The linear transformation $\qh=\qh(Q)$ implicit in (\ref{q4}) is not
invertible, that is $Q_{ijkl}=0$ does not imply $\qh_{ijkl}=0$, but one
will certainly be able to determine the non--pure gauge parts of 
$\qh_{ijkl}$ from this equation in terms of
$Q_{ijkl}$, and so in terms of the source. 
This will be seen more explicitly when
the solution $\gamma_{ij}$ is written in terms of these tensors.

To do this, the combinations of retarded moments which arise
in the solution 
(\ref{soltor4}) are obtained 
in terms of $m_i$, $\qh_{ijk}$ and $\qh_{ijkl}$.
For a pole--dipole--quadrupole source described by (\ref{gpdq}), this
solution can be written exactly as
\bgeq \gams_{ij}={2\over r}\tp_{ij} -{1\over r^2}
(2\tp_{ij:k}v^k+D{\tp_{ij:k}}^k) +{1\over r^3}(2\tp_{ij:kl}v^kv^l
-{\tp_{ij:k}}^k)\,.\label{pdqsol}\eneq
From (\ref{q4}) and (\ref{qdcm}), one has
\bgeq \tp_{ij:kl}=\qh_{ijpm}k^pk^mv_kv_l +\qh_{ijlp}k^pv_k +\qh_{ijkp}k^pv_l
+\qh_{ijkl}\,,\eneq
so that
\bgeq {\tp_{ij:k}}^k=-\qh_{ijkl}k^kk^l+2\qh_{ijkl}v^kk^l +{\qh_{ijk}}^k\,,
\eneq
and
\bgeq \tp_{ijkl}v^kv^l=\qh_{ijkl}k^kk^l-2\qh_{ijlp}v^lv^p 
+\qh_{ijkl}v^kv^l\,.\eneq
Using (\ref{q3}) and (\ref{q3sol}), one can write
\bgeq \tp_{ij:k} =-2\rdt_{ijl}k^lv_k -\ak R_{ijl}k^lv_k -R_{ijl}k^la_k
+k^lD\qh_{ijkl} -\qh_{ijl}k^lv_k -\qh_{ijk}\,,\eneq
and so
\bgeq \tp_{ij:k}v^k =2\rdt_{ijl}k^l+\ak R_{ijl}k^l 
+(\qh_{ijkl}k^l)^\cdot v^k +\qh_{ijl}k^l-\qh_{ijk}v^k\,,\eneq
and from (\ref{q3eq2}),
\bgeq m_{i:k}=\rdt_{ijk}k^k+\qh_{ijk}k^k\,.\eneq
Then a straightforward calculation gives
\begin{mathletters}\label{tinq}
\begin{eqnarray}
\tp_{ij}&=& {1\over 2}\qhtt_{ijkl}k^kk^l+{3\over 2}\ak \qht_{ijkl}k^kk^l
+{1\over 2}(\bk+3(\ak)^2)\qh_{ijkl}k^kk^l \nonumber\\
&&-\qht_{ijl}k^l-\ak \qh_{ijl}k^l -m_iv_j\,,\\
\nonumber\\
2\tp_{ij:k}v^k+D{\tp_{ij:k}}^k &=&
\qht_{ijkl}(\eta^{kl}+4v^kk^l-3k^kk^l)\nonumber\\
&&+\qh_{ijkl}(\ak(\eta^{kl}+6v^kk^l-6k^kk^l)+2a^kk^l)\nonumber\\
&&+2\qh_{ijl}(k^l-v^l)\,,\\
\nonumber\\
2\tp_{ij:kl}v^kv^l-{\tp_{ij:k}}^k&=&
\qh_{ijkl}(-\eta^{kl}+3k^kk^l-6v^kk^l+2v^kv^l)\,.\end{eqnarray}
\end{mathletters}

Thus from these expressions and (\ref{pdqsol}), it is seen that one can
express the solution in terms of arbitrary functions
of $u$, namely $m_i$ (which is in fact constant), $\qh_{ijk}(u)$ and 
$\qh_{ijkl}(u)$. One can go further than this, and show that these tensors 
act as potentials for the solution.
A direct calculation shows that
\begin{eqnarray}
\left({\qh^{ijl}\over r}\right)\!{_{,l}}&=&
-{1\over r}(\qht^{ijl}k_l+\ak\qh^{ijl}k_l)
+{1\over r^2}\qh^{ijl}(v_l-k_l)\,,\\
\left({\qh^{ijkl}\over r}\right)\!{_{,kl}}&=&
{1\over r}(\qhtt^{ijkl}+3\ak\qht^{ijkl}
+(\bk+3(\ak)^2)\qh^{ijkl})k_kk_l\nonumber\\
&&+{1\over r^2}\left( \qht^{ijkl}(-\eta_{kl}-4v_kk_l+3k_kk_l)\right.
\nonumber\\
&&+\left.
\qh^{ijkl}(-\ak(\eta_{kl}+6v_kk_l-6k_kk_l)-2a_kk_l)\Biggr)\right. \nonumber\\
&&+{1\over r^3}\left(\qh^{ijkl} (-\eta_{kl}-6v_kk_l+3k_kk_l+2v_kv_l)
\right)\,. \end{eqnarray}

Comparing these with (\ref{tinq}) and (\ref{pdqsol}) above, 
it has been shown that
\bgeq \gams^{ij}=-2{m^iv^j\over r} +2\left({\qh^{ijl}\over r}\right)\!{_{,l}}
+\left({\qh^{ijkl}\over r}\right)\!{_{,kl}}\,. \label{sbsol}\eneq
This form of the solution can be used to show that $\qh^{ijk}$ and 
$\qh^{ijkl}$ contain the same information as $Q^{ijk}$ and $Q^{ijkl}$.
To see this, notice that
\bgeq \left( {Q^{ijk}\over r} \right)\!{_{,k}}
=\left( {\qh^{ijk}\over r}\right)\!{_{,k}}
+\left( {\rdt^{ijk}\over r}\right)\!{_{,k}}\,,\eneq
and
\begin{eqnarray}
\left({Q^{ijkl}\over r}\right)\!{_{,kl}}&=&
\left({\qh^{ijkl}\over r}\right)\!{_{,kl}}
-2\left({R^{ijk}v^l\over r}\right)\!{_{,kl}}\nonumber\\
&=&
\left({\qh^{ijkl}\over r}\right)\!{_{,kl}}
-2\left({ {R^{ijk}}_{,l} v^l\over r}\right)\!{_{,k}}\nonumber\\
&=&
\left({\qh^{ijkl}\over r}\right)\!{_{,kl}}
-2\left({\rdt^{ijk}\over r}\right)\!{_{,k}}\,,
\end{eqnarray}
where $(v^i/r)_{,i}=0$ and $R_{ijk}=R_{ijk}(u,\ze,\zeb)$ have been used.
Thus 
\bgeq
2\left({Q^{ijk}\over r}\right)\!{_{,k}}
+\left({Q^{ijkl}\over r}\right)\!{_{,kl}}=
2\left({\qh^{ijk}\over r}\right)\!{_{,k}}
+\left({\qh^{ijkl}\over r}\right)\!{_{,kl}}\,,\eneq
and so (\ref{sbsol}) may equally be written without carats on the tensors
$\qh^{ijk}$ and $\qh^{ijkl}$. This validates the comments after 
Eq.(\ref{q4}); the kernel of the transformation only yields terms which do 
not contribute to $\gams^{ij}$, i.e. pure gauge terms.

Using the tensors $\qh^{ijk}$ and $\qh^{ijkl}$,
one could calculate what Janis and Newman \cite{JN} call the multipole
moments of the field (the coefficients of inverse powers of $r$ of the 
NP components of the curvature tensor). 
These would be given in terms of $v^i$, the $\qh$ tensors and known functions
of angle (from the null tetrad of \S 2).
This lengthy calculation
would result in combinations of our arbitrary functions of $u$ and such
spin--weighted spherical harmonics as are predicted by the general theory
\cite{NPC,me}.

The solution (\ref{sbsol})
is equivalent to those given by Sachs and Bergmann \cite{sb}
and Pirani \cite{pir} for `multipole particles', but has the advantage of
being derived from a realistic extended source rather than one confined 
to a time--like world--line. These authors make use of symmetric trace--free
tensors which are functions of the retarded time, and their solutions
contain infinite series of such tensors, corresponding to a particle
possessing an arbitrary number of multipole moments. 
It is clear then that for a general source for which arbitrarily many of
the retarded multipole moments (\ref{mlp}) are non-zero, the solution
(\ref{sbsol}) would contain an infinite series of tensors 
$\qh^{k_1\dots k_n}(u)$ which are arbitrary functions of $u$. Each of these
would be present at the $r^{-1}$ level in $\gamma_{ij}$, and would lead
to an infinite series of spin--weighted spherical
harmonics appearing in the fully decomposed expression for $\tp_{ij}$.
As seen in the previous section, in some situations, and especially
those involving a discussion of the radiation field, it is 
advantageous to use the retarded moments (\ref{mlp}). 

        
\section{Conclusions}
The similarity between the electromagnetic and linearized gravitational fields 
has been exploited in this paper to derive some results about the latter using
the Goldberg--Kerr approach to bound source fields. The results reflect those 
obtained by Hogan and Ellis \cite{HE} for the electromagnetic field.
Well established results (e.g. peeling-off) have been confirmed, relying on  
the use of the retarded multipole moments of the source. The advantage of
using these can be seen in, for example, Eq.(\ref{radflux}), where the total
radiated 4-momentum is determined by a single integral over the source, the
retarded monopole moment. The results of \S 5 show how this relates to 
analogous formulae of \cite{thorne}, which involve infinite series of
moments of the source. In addition, use of the retarded moments of the source
to describe the exterior field has made the consequences of the conservation
of the source easy to identify.

It comes as no surprise that a bound source linearized gravitational field
possesses a bound 4-momentum. The obvious question is: does this notion 
extend to the full non-linear gravitational field in the case of 
asymptotic flatness? The Landau-Lifschitz pseudotensor can be used to 
construct the 4-momentum for such fields 
(and the calculations would be carried out in an asymptotically Minkowskian
coordinate system, c.f. \S 20.2 of \cite{MTW}), and
indeed this 4-momentum conincides with the covariant (Bondi--Sachs)
constructions \cite{pp}. 
Hence one might 
expect that some or all of such fields do possess a bound 4-momentum, which
could be constructed in a manner similar to that above. 

\section{Conclusions}
The similarity between the electromagnetic and linearized gravitational fields 
has been exploited in this paper to derive some results about the latter using
the Goldberg--Kerr approach to bound source fields. The results reflect those 
obtained by Hogan and Ellis \cite{HE} for the electromagnetic field.
Well established results (e.g. peeling-off) have been confirmed, relying on  
the use of the retarded multipole moments of the source. The advantage of
using these can be seen in, for example, Eq.(\ref{radflux}), where the total
radiated 4-momentum is determined by a single integral over the source, the
retarded monopole moment. The results of \S 5 show how this relates to 
analogous formulae of \cite{thorne}, which involve infinite series of
moments of the source. In addition, use of the retarded moments of the source
to describe the exterior field has made the consequences of the conservation
of the source easy to identify.

It comes as no surprise that a bound source linearized gravitational field
possesses a bound 4-momentum. The obvious question is: does this notion 
extend to the full non-linear gravitational field in the case of 
asymptotic flatness? The Landau-Lifschitz pseudotensor can be used to 
construct the 4-momentum for such fields 
(and the calculations would be carried out in an asymptotically Minkowskian
coordinate system, c.f. \S 20.2 of \cite{MTW}), and
indeed this 4-momentum conincides with the covariant (Bondi--Sachs)
constructions \cite{pp}. 
Hence one might 
expect that some or all of such fields do possess a bound 4-momentum, which
could be constructed in a manner similar to that above. 

\appendix
\section{Second Derivatives of the Metric}
In order to obtain the second derivatives of the metric perturbation 
$\gamma_{ij}$ correct to $O(r^{-4})$, the solution (\ref{soltor4})
is written as
\be \gams_{ij}&=&{A_{ij}\over r}+{B_{ij}\over r^2}+{C_{ij}\over r^3}+
O(r^{-4})\,,\ee
where the coefficients may be read off from (\ref{soltor4}).
For convenience, the decomposition
\be B_{ij}&=&\alpha_{ij}+\beta_{ij}+\mu_{ij}\,,\ee
will be used, where
\be 
\alpha_{ij}&=&-4\tp_{ij:m}v^m\,,\\
\beta_{ij}&=&-2{\tpdt_{ij:m}}^m\,,\\
\mu_{ij}&=&-2\ak {\tp_{ij:m}}^m\,.
\ee
Then all the information required to construct the field (to $O(r^{-4})$) is 
contained in the second derivatives of the terms 
$A_{ij}$, $B_{ij}$, $\alpha_{ij}$, $\beta_{ij}$ and $\mu_{ij}$.
These derivatives are obtained using results given in \S\S 1-3.
It is found that
\be 
\left({A_{ij}\over r}\right)\!{_{,kl}}=
&\displaystyle\frac{4}{r}&\left((D+\ak )D\tp_{ij}\right)k_kk_l\nn\\
-&\displaystyle\frac{4}{r^2}&\left\{2(D+\ak )D\tp_{ij:m}
(\delta^m_{(k}+v^mk_{(k})k_{l)}
+D\tp_{ij:m}(a^m+\ak  v^m)k_lk_k \right.\nn\\
&&\left.D\tp_{ij}(\eta_{kl}+4v_{(k}k_{l)}-3k_kk_l)+\ak \tp_{ij}
(2v_{(k}k_{l)}-3k_kk_l)+2\tp_{ij}(a_{(k}k_{l)}) \right\}\nn\\
+&\displaystyle\frac{4}{r^3}&
\left\{ 
\tp_{ij}(-\eta_{kl}+2v_kv_l-6v_{(k}k_{l)}+3k_kk_l)
+v^mD\tp_{ij:m}(\eta_{kl}+2v_{(k}k_{l)}-k_kk_l)\right.\nn\\
&&+2\tp_{ij:m}a_{(k}(\delta^m_{l)}+v^mk_{l)}) 
+ 2(2D+\ak )\tp_{ij:m}(\delta^m_{(l}+v^mk_{(l})(v_{k)}-k_{k)})\nn\\
&&\left.
+(D+\ak )D\tp_{ij:mp}(\delta^m_k+v^mk_k)(\delta^p_l+v^pk_l) \right\}
\label{first}\,,
\ee
\be
\left({\alpha_{ij}\over r^2}\right)\!{_{,kl}}=
-&\displaystyle\frac{4}{r^2}&\left\{
(D+2\ak )(D+\ak )(v^m\tp_{ij:m})
\right\} k_kk_l \nn\\
+&\displaystyle\frac{4}{r^3}&\left\{
(v^m\tdot_{ij:m}+a^m\tp_{ij:m})(\eta_{kl}+3v_kk_l+3v_lk_k-5k_kk_l)\right.
\nn\\
&&+2v^m\tp_{ij:m}(a_kk_l+a_lk_k+\ak(\eta_{kl}+4v_kk_l+4v_lk_k-8k_kk_l))\nn\\
&&+(-(D+2\ak)\tp_{ij:p}+v^m(D+2\ak)D_{ij:mp}+a^m\tp_{ij:mp})\times\nn\\
&&((\delta^p_l+v^pk_l)k_k+(\delta^p_k+v^pk_k)k_l)\nn\\
&&\left.+(a^p+\ak v^p)(-\tp_{ij:p}+D\tp_{ij:mp})k_kk_l\right\}+O(r^{-4})\,.
\label{second}
\ee
\be
\left( {\beta_{ij}\over r^2}\right)\!{_{,kl}}=
-&\displaystyle\frac{2}{r^2}&(D+2\ak )(D+\ak )
\left.{\tdot_{ij:m}}\right.^m k_kk_l\nn\\
+&\displaystyle\frac{2}{r^3}&
\left\{ \left.\tdtt_{ij:m}\right.^m(\eta_{kl}+3v_kk_l+3v_lk_k-5k_kk_l)\right.
\nn\\
&&+2\left.{\tdot_{ij:m}}\right.^m
(a_kk_l+a_lk_k+\ak(\eta_{kl}+4v_kk_l+4v_lk_k-8k_kk_l))
\nn\\
&&+(2(D+2\ak)D(v^m\tp_{ij:mp})+(D+2\ak)D^2{\tp_{ij:pm}}^m)\times\nn\\
&&((\delta^p_l+v^pk_l)k_k+(\delta^p_k+v^pk_k)k_l)\nn\\
&&\left.+(a^p+\ak v^p)(2v^mD\tp_{ij:mp}+2a^m\tp_{ij:mp}
+D^2\left.{\tp_{ij:pm}}\right.^m)k_kk_l
\right\}+O(r^{-4})\,.\label{third}
\ee

\be
\left({\mu_{ij}\over r^2}\right)\!{_{,kl}}=
-&\displaystyle\frac{2}{r^2}&(D+2\ak )(D+\ak )(\ak {\tp_{ij:m}}^m)k_kk_l\nn\\
+&\displaystyle\frac{2}{r^3}&\left\{
2(\bk +3\ak ^2)(2v^m\tp_{ij:mp}+D{\tp_{ij:pm}}^m)
(\delta^p_{(l}+v^pk_{(l})k_{k)}\right.\nn\\
&&+2\ak (2v^mD\tp_{ij:mp}+2a^m\tp_{ij:mp}+D^2{\tp_{ij:pm}}^m)
(\delta^p_{(l}+v^pk_{(l})k_{k)}\nn\\
&&\left.+\ak(2v^m\tp_{ij:mp}+D{\tp_{ij:pm}}^m)(a^p+\ak v^p)k_kk_l\right\}
+O(r^{-4})\,.
\label{fourth}
\ee
\be
\left({C_{ij}\over r^3}\right)\!{_{,kl}}=
&\displaystyle\frac{1}{r^3}&(D+3\ak )(D+2\ak )C_{ij}k_kk_l + O(r^{-4})\,.
\label{last}
\ee

\newcommand{\jmp}{J.~Math.~Phys.}

\end{document}